\documentclass[a4paper]{jpconf}
\usepackage{graphicx}
\usepackage{amsmath,amsfonts,amssymb,color}
\newcommand{\bayreuth}{$^1$ Theoretische Physik III, Universit\"at Bayreuth, 95440 Bayreuth, Germany}

\newcommand{\buenosaires}{$^2$ Departamento de F\'isica and IFIBA, FCEN, Universidad de Buenos Aires, 
Ciudad Universitaria, Pab.\ I, C1428EHA Buenos Aires, Argentina}

\begin{document}

\title{Relaxation and coherent oscillations in the  spin dynamics of 
II-VI diluted magnetic quantum wells}

\author{F.\ Ungar$^1$, M.\ Cygorek$^1$, P.\ I.\ Tamborenea$^{1,2}$ and V.\ M.\ Axt$^1$} 
\address{\bayreuth}
\address{\buenosaires}

\date{\today}

\begin{abstract}
We study theoretically the ultrafast spin dynamics of II-VI diluted magnetic 
quantum wells in the presence of spin-orbit interaction.
We extend a recent study where it was shown  that the spin-orbit interaction and the 
exchange sd coupling in bulk and quantum wells can compete resulting in 
qualitatively new dynamics when they act simultaneously.
We concentrate on Hg$_{1-x-y}$Mn$_x$Cd$_y$Te quantum wells, which have a highly tunable
Rashba spin-orbit coupling.
Our calculations use a recently developed formalism which incorporates 
electronic correlations originating from the exchange $sd$-coupling.
We find that the dependence of electronic spin oscillations on the excess energy changes 
qualitatively depending on whether or not the spin-orbit interaction dominates or is of 
comparable strength with the sd interaction.
\end{abstract}


Ultrafast spin dynamics in semiconductors is attracting nowadays a great deal of attention.
In a recent article we explored the interplay between the exchange sd interaction (EXI)
and the spin-orbit interaction (SOI) in II-VI diluted magnetic semiconductors 
(DMS) \cite{ung-cyg-tam-axt}.
In that study we found that the EXI and the SOI
can be tuned to overrun each other or to compete on an equal level in realistic
bulk and quantum well systems.
Importantly, we found that the inclusion of the SOI  introduces
oscillations in the spin dynamics which are completely absent when only the
EXI  is relevant.
In the present article, we characterize systematically the decay and the periods of 
oscillations seen in the spin dynamics in quantum wells, 
as functions of the excess energy of the electron population in the conduction 
band (mean energy of a Gaussian occupation of spin-polarized photoexcited electrons).
We employ a microscopic density-matrix theory that models on a quantum-kinetic level the spin
dynamics, taking into account the exchange-induced correlations and the localized character 
of the Mn spins \cite{thu-axt, cyg-axt-15}.
For the sake of brevity we shall only sketch the model here and refer the reader to 
Refs.\ \cite{ung-cyg-tam-axt} and \cite{cyg-axt-15}
for a complete description.

We consider conduction band electrons in 
Mn-doped II-VI DMS coming from low-intensity optical excitations.
We work in the regime of electron densities $n_e$ 
much lower than the Mn density $n_{\text{Mn}}$, in which the 
Mn spin variables are nearly stationary 
\cite{cyg-axt-15, cyg-axt, thu-cyg-axt-kuh-b, thu-cyg-axt-kuh-a}.
%
%
The spin dynamics is described by 
$\langle \mathbf{s}_{\mathbf{k}}^{\perp}\rangle(t)$ and
$\langle s_{\mathbf{k}}^{\parallel} \rangle(t)$,
the mean electronic spin components, perpendicular and parallel
to the Mn magnetization, respectively, corresponding to the conduction-band 
state $\mathbf{k}$.

\begin{figure}[t]
\centerline{\includegraphics[width=6.2in]{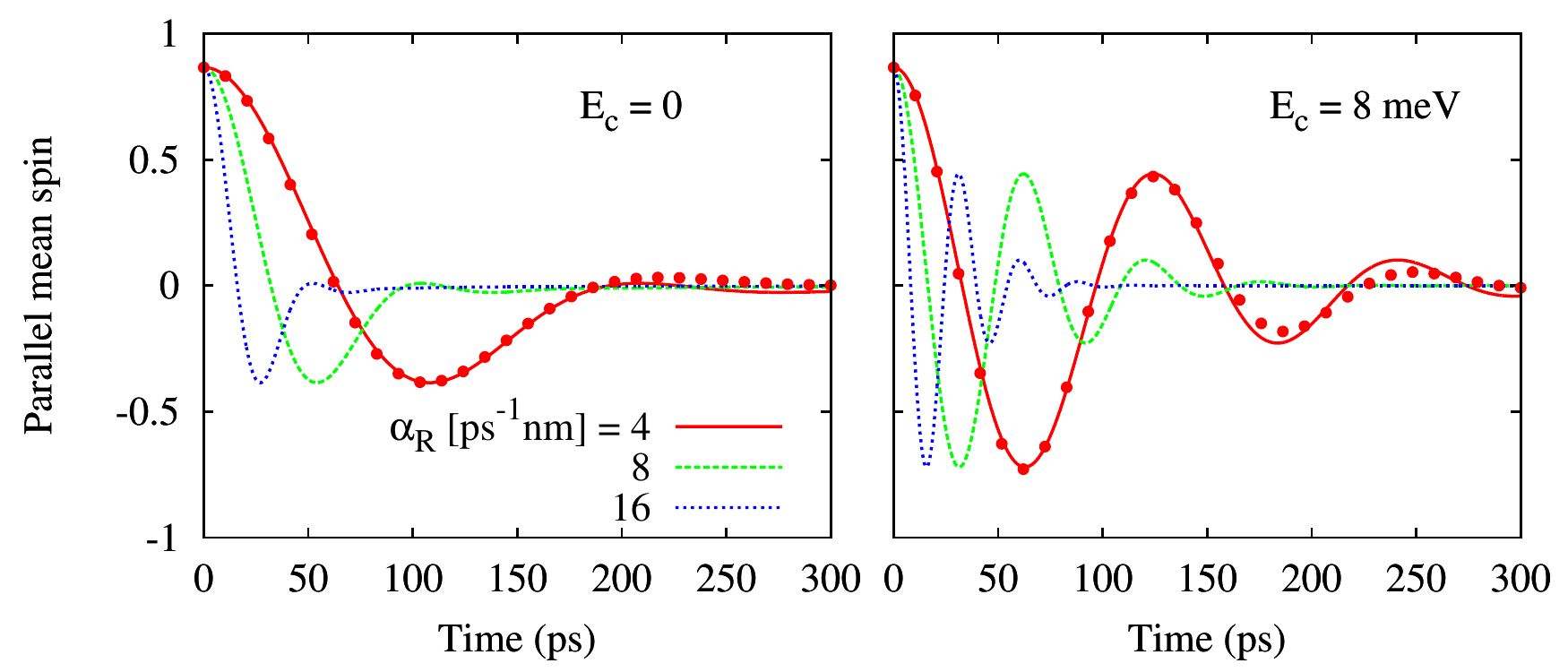}}
\caption{
  Electron spin dynamics in a Hg$_{1-x-y}$Mn$_x$Cd$_y$Te quantum well with Rashba interaction
  and no exchange sd coupling for a 
  Gaussian electron distribution centered at $E_c$.
  Lines: simulations for different values of $\alpha_R$ expressed 
  in units of $\text{ps}^{-1} \text{nm}$.
  Red dots: simple fits to the first oscillation of the curves with 
  $\alpha_R = 4 \, \text{ps}^{-1} \text{nm}$ (see text for details).
}
\label{fig:QW_rashba}
\end{figure}

In this study we concentrate on Hg$_{1-x-y}$Mn$_{x}$Cd$_{y}$Te quantum wells, 
since this alloy offers great flexibility in the control of the Rashba SOI.
This control is achieved thanks to the strong dependence of the band gap on
the doping fraction $x+y$ \cite{kos}.
With this material, Rashba coefficients of the order of
$\alpha_R \approx 10 \, \text{ps}^{-1} \text{nm}$
can be obtained for realistic quantum well specifications \cite{ung-cyg-tam-axt}.
Throughout the paper we shall assume the Mn magnetization
to be perpendicular to the quantum well. 
Furthermore, we take a Gaussian electron occupation centered at an
energy $E_c$ above the conduction-band edge with a standard
deviation of $E_s=$~3~meV and initial spin-polarization rotated 30$^{\circ}$ 
with respect to the Mn magnetization.

\begin{figure}[t]
\begin{minipage}{18pc}
\includegraphics[width=18pc]{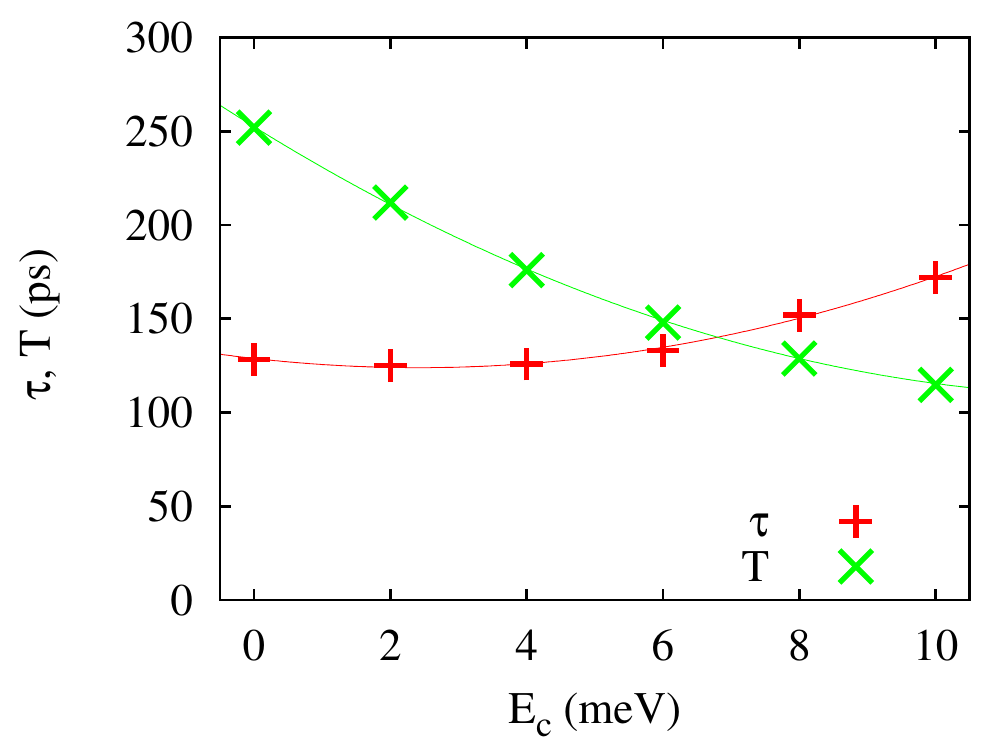}
\caption{
  \label{fig:tau_T_vs_Ec}
  Gaussian electron spin relaxation time, $\tau$, and period of the oscillations, $T$, 
  in a Hg$_{1-x-y}$Mn$_x$Cd$_y$Te quantum well with Rashba interaction 
  $\alpha_R = 4 \, \text{ps}^{-1} \text{nm}$ and no exchange sd coupling 
  as a function of the excess energy $E_{c}$. 
  Symbols: full calculation; lines: parabolic fit.
  }
\end{minipage}\hspace{2pc}
\begin{minipage}{18pc}
\includegraphics[width=18pc]{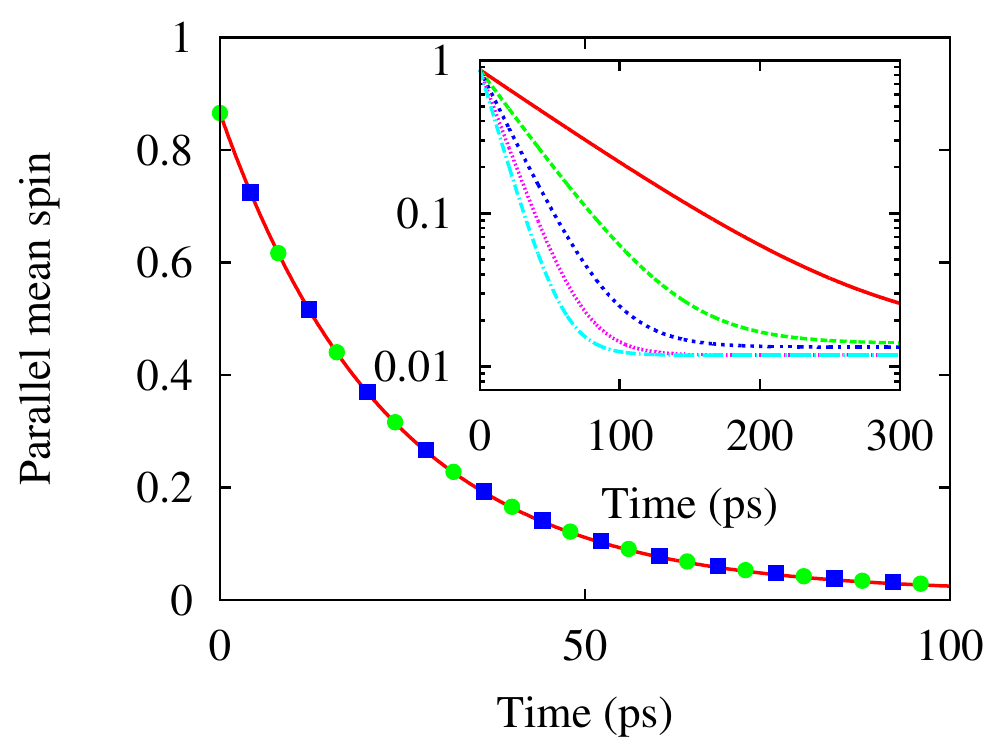}
\caption{
  \label{fig:only_sd}
   $\langle s^\|(t)\rangle$  with exchange sd and no Rashba interaction.
  $x_{\text{Mn}}=$ 0.3\%, $S = 0.1$, $E_c=0$ (red line),  
  $4 \, \text{meV}$ (green circles), $8 \, \text{meV}$ (blue squares).
  Inset: $E_c=0$,  $S = 0.1$, various $x_{\text{Mn}}$:
  0.1\% (red), 0.2\% (green), 0.3\% (blue), 0.4\% (magenta), 0.5\% (cyan).}
\end{minipage} 
\end{figure}

It is instructive to examine first the spin dynamics resulting only from the Rashba 
spin-orbit interaction (i.e., no exchange sd coupling is accounted for). 
In this case the Mn magnetization does not enter the dynamics. 
For later comparison we used, however, the above described initial condition where
the direction of the initial electronic spin is related to direction of the 
Mn magnetization.
Figure \ref{fig:QW_rashba} shows the time evolution of the summed parallel spin 
component, 
$\langle s^{\parallel} \rangle(t) = 
 \sum_{\mathbf{k}} \langle s_{\mathbf{k}}^{\parallel} \rangle(t)$, for
three different values of the Rashba coupling constant $\alpha_R$.
%
The cases $E_c=0$ (Gaussian occupation centered at the band edge) and $E_c= 8 \, \text{meV}$
are shown.
We see that the Rashba interaction produces well-defined oscillations and decay.
Note that without the EXI, the time evolution of the total spin is given by coherent 
precessions of individual electron spins around the $\mathbf k$-dependent magnetic 
Rashba field, which are collectively responsible for the decay.
Red dots in both panels of Fig.\ \ref{fig:QW_rashba} are fits to the initial oscillations
of the $\alpha_R = 4 \, \text{ps}^{-1} \text{nm}$ evolutions, done with
a function $f(t) \propto \exp{[-(t/\tau)^2]} \cos{(2\pi t/T)}$.
For a Gaussian electron distribution with spins pointing in the growth direction, 
we find from Eq.\ (17) of Ref.~\cite{ung-cyg-tam-axt}:
\begin{eqnarray}
\langle s^\|(t)\rangle = 
   C \int_{0}^{\infty} dk\; k \,
   \exp\left[
           -\frac{\big(\hbar^2 k^2-2m^* E_c\big)^2} 
                 {(2 m^* E_s)^2}
       \right]
       \cos(2\alpha_R k t),
\label{s-para}
\end{eqnarray}
where $m^*$ is the effective mass and $C$ is a constant determined by the
initial value of the total spin. 
The integral in Eq.~(\ref{s-para}) is close to the (half-sided) Fourier transform of a function
with a single central peak indicating in time regime a damped oscillation with roughly the
peak frequency. 
An initially exponential decay of $\langle s^\|(t)\rangle$ would require
a Lorentzian decay in the energy domain. 
However, the function in Eq.~(\ref{s-para}) decays much faster for large $k$ explaining 
why the initial behavior of $\langle s^\|(t)\rangle$ is much better approximated by 
a Gaussian than by an exponential.
Indeed, Fig.\ \ref{fig:QW_rashba} reveals that 
the Gaussian fit is almost perfect at early times but worsens somewhat later.
Applying the Gaussian fit to a number of different cases we find that for given $E_c$  
both $\tau \propto \alpha_R^{-1}$ and $T \propto \alpha_R^{-1}$
hold to a very good approximation.
A similar behavior is observed for 
$|\langle \mathbf{s}^{\perp}\rangle(t)| = 
|\sum_{\mathbf{k}} \langle \mathbf{s}_{\mathbf{k}}^{\perp} \rangle(t)|$,
whose initial evolution can be well fitted with 
$f(t) \propto \exp{[-(t/\tau)^2]} \cos{(2\pi t/T)} + 1$,
with the same values of $\tau$ and $T$ as for $\langle s^{\parallel} \rangle(t)$.
We found that $\tau$ and $T$ can be precisely fitted as functions of $E_c$
with parabolas, as seen in Fig.\ \ref{fig:tau_T_vs_Ec}.
The decrease of $T$ and the increase of $\tau$ with rising $E_{c}$ reflect the fact
that the effective Rashba field is $k$ dependent becoming stronger for
larger $k$.

Let us now look at the spin dynamics under the influence of the EXI only 
(no Rashba SOI).
For the EXI coupling constant we take the value 
$J_{\text{sd}} = 26.8 \, \text{meV}\,\text{nm}^3$ \cite{ung-cyg-tam-axt}.
The effects of the EXI  on the carrier spins can be controlled via the 
Mn concentration, $x_{\text{Mn}}$, and the initial net Mn magnetization, 
$S = |\langle \mathbf{S} \rangle|$ \cite{ung-cyg-tam-axt}.
The dynamics of the electron spin component parallel to the Mn magnetization
is shown in Fig.~\ref{fig:only_sd}. 
As found previously, it is approximately described by an exponential decay 
to an in general non-zero equilibrium value with a decay time 
[cf. Eq.(19) of Ref.~\cite{cyg-axt-15}]
$\tau^\| = (J_{sd}^2 n_{Mn} m^* / \hbar^2 d) \langle S^2- {S^\|}^2 \rangle,$
where $d$ is the width of the quantum well and $\langle S^2-{S^\|}^2\rangle$
is a second moment of the spin-$\frac 52$ Mn system.
In particular, $\tau^\|$ is linear in $x_{\text{Mn}}$  and independent of 
the excess energy $E_c$. Note that no oscillations appear in the
time evolution of the parallel spin component when only the EXI
is present.
The perpendicular component decays to zero  with a slightly different 
rate while it precesses around the Mn magnetization \cite{cyg-axt-15}.

\begin{figure}[t]
\begin{minipage}{18pc}
\includegraphics[width=18pc]{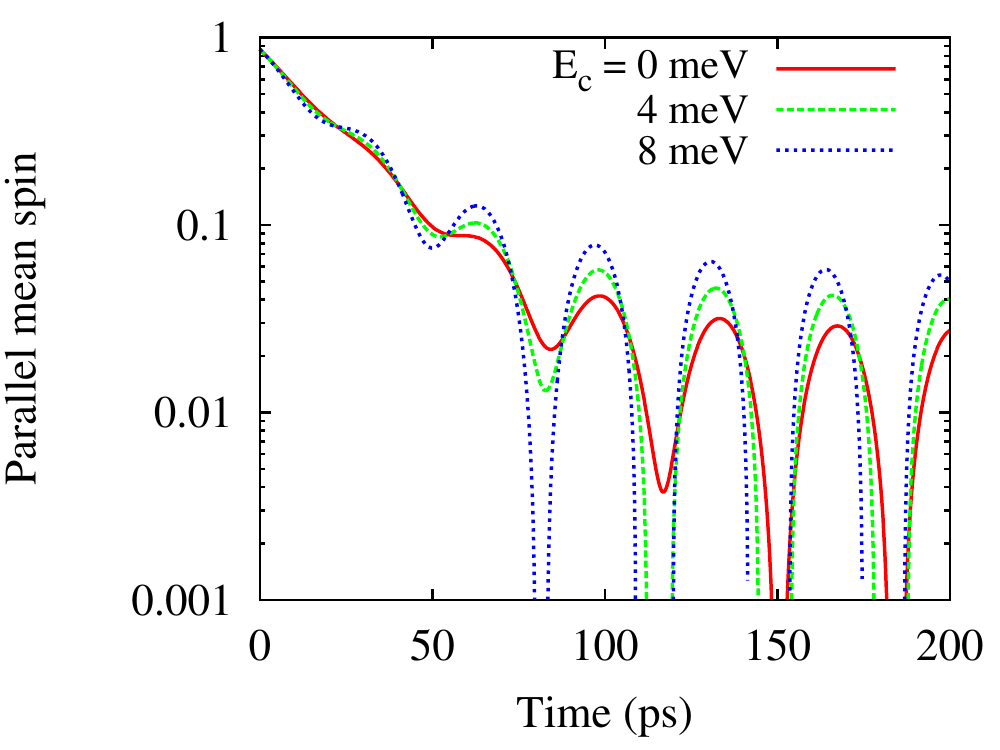}
\caption{
  \label{fig:strong_sd_weak_R}
  Evolution of the mean parallel electron spin under exchange sd and Rashba interactions.
  Parameters: $x_{\text{Mn}} = 0.3\%$, $S = 0.1$, 
  $\alpha_R \approx 4 \, \text{ps}^{-1} \text{nm}$, and three values of $E_c$.
  }
\end{minipage}\hspace{2pc}
\begin{minipage}{18pc}
\includegraphics[width=18pc]{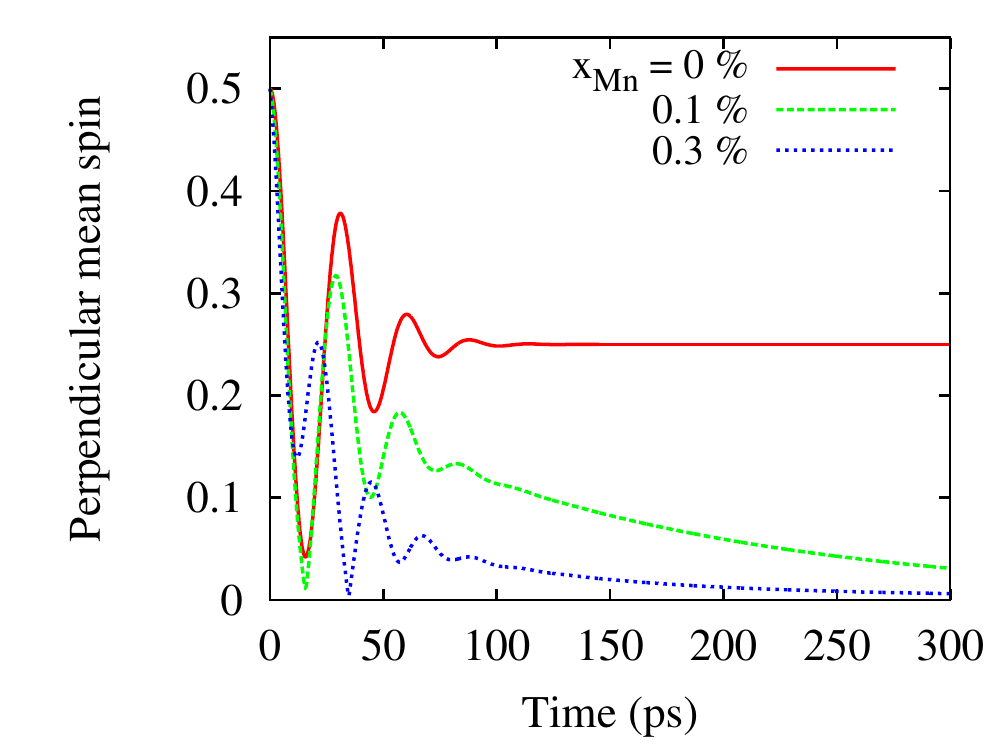}
\caption{
  \label{fig:weak_sd_strong_R}
  Evolution of the mean perpendicular electron spin component under exchange sd and Rashba interactions.
  Parameters: $\alpha_R = 16 \, \text{ps}^{-1} \text{nm}$, $E_c= 8 \, \text{meV}$, 
  $S = 0.1$, and three values of $x_{\text{Mn}}$.
  }
\end{minipage} 
\end{figure}

We now come to the combined effects of the Rashba SOI and EXI.
Figure \ref{fig:strong_sd_weak_R} shows the spin dynamics for $x_{\text{Mn}} = 0.3\%$, 
$S = 0.1$, $\alpha_R \approx 4 \, \text{ps}^{-1} \text{nm}$, 
and three different values of $E_c$.
Note the semilog scale chosen to better visualize the long time behavior.
This set of parameters defines a ``strong sd'' and ``weak Rashba'' situation.
Accordingly, the initial decay is exponential (not Gaussian like in Fig.\ \ref{fig:QW_rashba}), 
with an $E_c$-independent decay rate, like in Fig.\ \ref{fig:only_sd}.
However, at later times fairly regular oscillations appear, with essentially constant, $E_c$-dependent 
amplitude, due to the presence of the Rashba interaction.
We note that the frequency of the oscillations depends only slightly on $E_c$, becoming higher
for higher $E_c$, which indicates a slight dependence on the Rashba mechanism.
The frequency of the oscillations (period of about 35 ps) is close to the precession frequency 
about the net Mn magnetic field, which is independent of $E_c$.
The oscillations do not decay because all electrons precess with nearly the same frequency 
governed mainly by the Mn magnetization and thus do not dephase.
The fact that the amplitude of the oscillations does not decay with time and 
the near independence of the period on $E_c$ distinguish qualitatively 
these oscillations from the ones observed with Rashba coupling alone.
Also note that again there is a saturation value different from zero as seen above in the sd-only
case.
However, a new feature produced by the presence of the Rashba interactions is that the perpendicular 
spin component does not go to zero as in the sd-only evolution (not shown for brevity).

Finally, Fig.\ \ref{fig:weak_sd_strong_R} shows the time evolution of the mean perpendicular 
spin component under EXI and SOI for fixed Rashba constant
$\alpha_R = 16 \, \text{ps}^{-1} \text{nm}$, $E_c= 8 \, \text{meV}$, 
and three different values of $x_{\text{Mn}}$.
This figure shows the effect of an increasing EXI coupling in the presence of a strong 
Rashba coupling on the perpendicular spin component.
We see that as the Mn concentration increases, 
starting from a Rashba-only situation in which the equilibrium value 
$\langle s_{\mathbf{k}}^{\perp}\rangle$
is half its initial value \cite{ung-cyg-tam-axt}, 
a decay to zero sets in.
At the same time, the frequency of the oscillations increases and their amplitude goes down.

In conclusion, we have studied theoretically the effects of the Rashba spin-orbit interaction 
in II-VI diluted-magnetic-semiconductor quantum wells.
We characterized the dependence of the spin dynamics on the excess energy of a Gaussian
population of electrons in the conduction band.
Our findings provide qualitative signatures that could aid experimentalists in distinguishing
the relative importance of spin-orbit and exchange interactions in DMS quantum wells.

We gratefully acknowledge the financial support of the Deutsche Forschungsgemeinschaft
(grant No.\ AX17/9-1), 
the Universidad de Buenos Aires (UBACyT 2011-2014 No.\ 20020100100741), 
and CONICET (PIP 11220110100091).

\section*{References}

\medskip

\bibliography{confpapbib}

\end{document}